\documentclass[conference]{IEEEtran}
\ifCLASSINFOpdf
\else
\fi

\usepackage{pdfpages}
\usepackage{graphicx}

\usepackage{bm}
\usepackage{enumerate}
\usepackage{amssymb}
\usepackage{float}
\usepackage{multicol}
\usepackage[keeplastbox]{flushend}
\usepackage{color}
\usepackage{amsmath}
\usepackage{algorithmic}
 \usepackage{algorithm}
\usepackage{array}
\ifCLASSOPTIONcompsoc
  \usepackage[caption=false,font=normalsize,labelfont=sf,textfont=sf]{subfig}
\else
  \usepackage[caption=false,font=footnotesize]{subfig}
\fi
\begin{document}
\title{A Genetic Algorithm-based Beamforming Approach for Delay-constrained Networks }

% \author{\IEEEauthorblockN{Michael Shell}
% \IEEEauthorblockA{School of Electrical and\\Computer Engineering\\
% Georgia Institute of Technology\\
% Atlanta, Georgia 30332--0250\\
% Email: http://www.michaelshell.org/contact.html}
% \and
% \IEEEauthorblockN{Homer Simpson}
% \IEEEauthorblockA{Twentieth Century Fox\\
% Email: homer@thesimpsons.com}
% \and
% \IEEEauthorblockN{James Kirk\\ and Montgomery Scott}
% \IEEEauthorblockA{Starfleet Academy\\
% San Francisco, California 96678--2391\\
% Telephone: (800) 555--1212\\


\author{\IEEEauthorblockN{Michael Shell\IEEEauthorrefmark{1},
Homer Simpson\IEEEauthorrefmark{2},
James Kirk\IEEEauthorrefmark{3}, 
Montgomery Scott\IEEEauthorrefmark{3} and
Eldon Tyrell\IEEEauthorrefmark{4}}
\IEEEauthorblockA{\IEEEauthorrefmark{1}School of Electrical and Computer Engineering\\
Georgia Institute of Technology,
Atlanta, Georgia 30332--0250\\ Email: see http://www.michaelshell.org/contact.html}
\IEEEauthorblockA{\IEEEauthorrefmark{2}Twentieth Century Fox, Springfield, USA\\
Email: homer@thesimpsons.com}
\IEEEauthorblockA{\IEEEauthorrefmark{3}Starfleet Academy, San Francisco, California 96678-2391\\
Telephone: (800) 555--1212, Fax: (888) 555--1212}
\IEEEauthorblockA{\IEEEauthorrefmark{4}Tyrell Inc., 123 Replicant Street, Los Angeles, California 90210--4321}}

\author{\IEEEauthorblockN{Hao Guo, Behrooz Makki, Tommy Svensson}
\IEEEauthorblockA{Department of Signals and Systems, Chalmers University of Technology, Gothenburg, Sweden
    \\ghao@student.chalmers.se, \{behrooz.makki, tommy.svensson\}@chalmers.se}}


\maketitle

\begin{abstract}
In this paper, we study the performance of initial access beamforming schemes in the cases with large but finite number of transmit antennas and users. Particularly, we develop an efficient beamforming scheme using genetic algorithms. Moreover, taking the millimeter wave communication characteristics and different metrics into account, we investigate the effect of various parameters such as number of antennas/receivers, beamforming resolution as well as hardware impairments on the system performance. As shown, our proposed algorithm is generic in the sense that it can be effectively applied with different channel models, metrics and beamforming methods. Also, our results indicate that the proposed scheme can reach  (almost) the same end-to-end throughput as the exhaustive search-based optimal approach with considerably less implementation complexity. 
\end{abstract}


% \fi
\IEEEpeerreviewmaketitle

\section{Introduction}
Developing key technical components and concepts for millimeter wave (MMW) communications in the range of 6-100 GHz is of interest for 5G. Different works have estimated/measured the channel characteristics in such MMW frequency bands [1]-[4]. The use of such high frequencies for mobile communications is challenging but[5].

Due to peak power limitation and high path loss in MMW communications, there is a need for directional transmissions. Fortunately, the physical size of antennas at MMW frequency bands is small so that it is possible to use large antenna arrays and perform beamforming [3][4]. For typical wireless systems, beamforming is performed by employing precoding with channel state information (CSI) feedback or estimation after the control link is established. However, even with the extended coverage from beamforming, the coverage range for MMW frequencies is typically small due to high pathloss. As a result, we need to employ beamforming also on initial access (IA) channels. This calls for the need to design novel IA procedures. 

During the MMW initial access procedure, beamforming is different from the conventional one because it is hard to acquire CSI. Different works have been recently presented on both physical architecture and procedural algorithm to solve the problem (see Section II for literature review).  However, in these works either the initial access algorithms are designed for specific metrics, channel models, and precoding schemes or their implementation complexity grows significantly with the number of antennas/users. Moreover, the running delay of the algorithm is an important issue which has been rarely considered in the performance evaluations. 

In this paper, we study the performance of large-but-finite multiple-input-multiple-output (MIMO) MMW networks using codebook-based beamforming. The contributions of the paper are two-fold. First, we propose an efficient genetic algorithm (GA)-based approach for initial access beamforming. With the proposed algorithm, the appropriate beamforming matrix is selected from a set of predefined matrices such that the network performance is optimized. As we show, our proposed scheme is generic in the sense that it can be implemented in the cases with different channel models, beamforming methods as well as optimization metrics. Second, we evaluate the performance of the beamforning-based MIMO networks for different parameters such as hardware impairments, different channel models, beamforming resolution and number of antennas/receivers. 

For the simulation results, we consider the end-to-end throughput, the end-to-end service outage-constrained throughput as well as the service outage probability. In this way, we take the algorithm running time into account and, as opposed to conventional iterative schemes, the system performance is not necessarily improved in successive iterations. Instead, as shown via simulations, the maximum throughput is achieved with few iterations, i.e., by picking up a suboptimal beamforming approach, and using the remaining time for information transmission. 

The simulation results show that 1) the proposed scheme can reach (almost) the same performance as in the exhaustive search-based scheme with considerably less implementation complexity. Moreover, 2) non-ideal power amplifiers (PAs) affect the system performance significantly and should be carefully considered/compensated in the network design, and 3) the network throughput increases almost linearly with the number of codebook vectors. 4) The proposed algorithm is effectively applicable for various convex and non-convex performance metrics. 5) In general, the users service outage constraints affect the end-to-end throughput remarkably, while the effect of the constraint decreases at high signal-to-noise ratios (SNRs). 6) In practice, taking the algorithm running delay into account, the maximum end-to-end throughput is reached by dedicating a small fraction of the packet period to finding the suboptimal beamforming solution and using the rest of the packet for data transmission.

% This demo file is intended to serve as a ``starter file''
% for IEEE conference papers produced under \LaTeX\ using
% IEEEtran.cls version 1.8b and later.
% % You must have at least 2 lines in the paragraph with the drop letter
% % (should never be an issue)
% I wish you the best of success.

 

% \subsection{Subsection Heading Here}
% Subsection text here.

% \subsubsection{Subsubsection Heading Here}
% Subsubsection text here.

\section{Literature Review}
In this section, we review the recent results on initial access. The readers mainly interested in the technical discussions can skip this part and go to Sections III-V where we present the system model, our proposed algorithm and the simulation results, respectively. Beamforming techniques at MMW frequencies have been widely investigated and led to standard developments such as IEEE 802.15.3c (TG3c)[6], IEEE 802.11ad (TGad)[7] and ECMA-387[8]. In wireless local area networks (WLANs), a one-sided beam search strategy using a beamforming codebook has been employed to establish initial alignment between large array antennas[9]. 

For mobile communication systems, on the other hand, there are few works on initial access beamforming. In general, most of the presented works are based on multi-level/greedy search algorithms and utilize the sparse nature of the MMW channel. Several issues for initial access beamforming in MMW frequencies are presented in [10] and a fast-discovery hierarchical search method is proposed. Moreover, [11] designs a novel greedy-geometric algorithm to synthesize antenna patterns featuring desired beamwidth. In [12], a survey of several recently proposed IA techniques is provided. Then,  [13] performs  IA for clustered MMW small cells  with a power-delay-profile-based approach to reduce the IA set up time. Also, [14] shows the significant benefits of using low-resolution fully digital architectures during IA, in comparison to single stream analog beamforming.

In MMW multiuser multiple-input-single-output (MISO) downlink systems, an opportunistic random beamforming technique is provided in [15]. Also, [16] develops low-complexity algorithms for optimizing the choice of beamforming directions, premised on the sparse multipath structure of the MMW channel. Then, [17] studies a low-complexity beam selection method by designing low-cost analog beamformers. This beam selection method can be carried out without explicit channel estimation. A directional cell discovery method is proposed in [18] where the BS periodically transmits synchronization signals to scan the whole angular space in time-varying random directions. In [19], MMW precoder design is formulated as a sparsity-constrained signal recovery problem, and an algorithmic solution with orthogonal matching pursuit is proposed. Finally, [20] proposes a hybrid precoding algorithm based on a low training overhead channel estimation method to overcome the hardware constraints in MMW analog-only beamforming systems.

Considering codebook-based beamforming, a broadcast-based solution for MMW systems is proposed in [21], where limited feedback-type directional codebooks are used for the beamforming procedure. Moreover, [22] studies concurrent beamforming issues for achieving high capacity in indoor MMW networks. In [23], an efficient beam alignment technique is designed which uses adaptive subspace sampling and hierarchical beam codebooks. With pre-specified beam codebooks, [24] proposes a Rosenbrock numerical algorithm to accelerate the beam-switch process which is modeled as a 2-D plane optimization problem. Also, [25] adopts the discrete Fourier transform (DFT)-based codebooks and proposes an efficient iterative antenna vector training algorithm. Finally, [26] provides a codebook-based beamforming scheme with multi-level training and level-adaptive antenna selection, which can be used for MIMO orthogonal frequency division multiplexing (OFDM) systems in MMW wireless personal area networks (WPANs).

There are previous works using the GA-based selection approach. For instance, [27] elaborates on the performance of scheduling in the return-link of a multi-beam satellite system. Moreover, [28] uses a genetic algorithm to achieve a near-optimal array gain in all directions during codebook-based beamforming.

\section{System Model}
\begin{figure}[!t]
\centering
\includegraphics[width=3in]{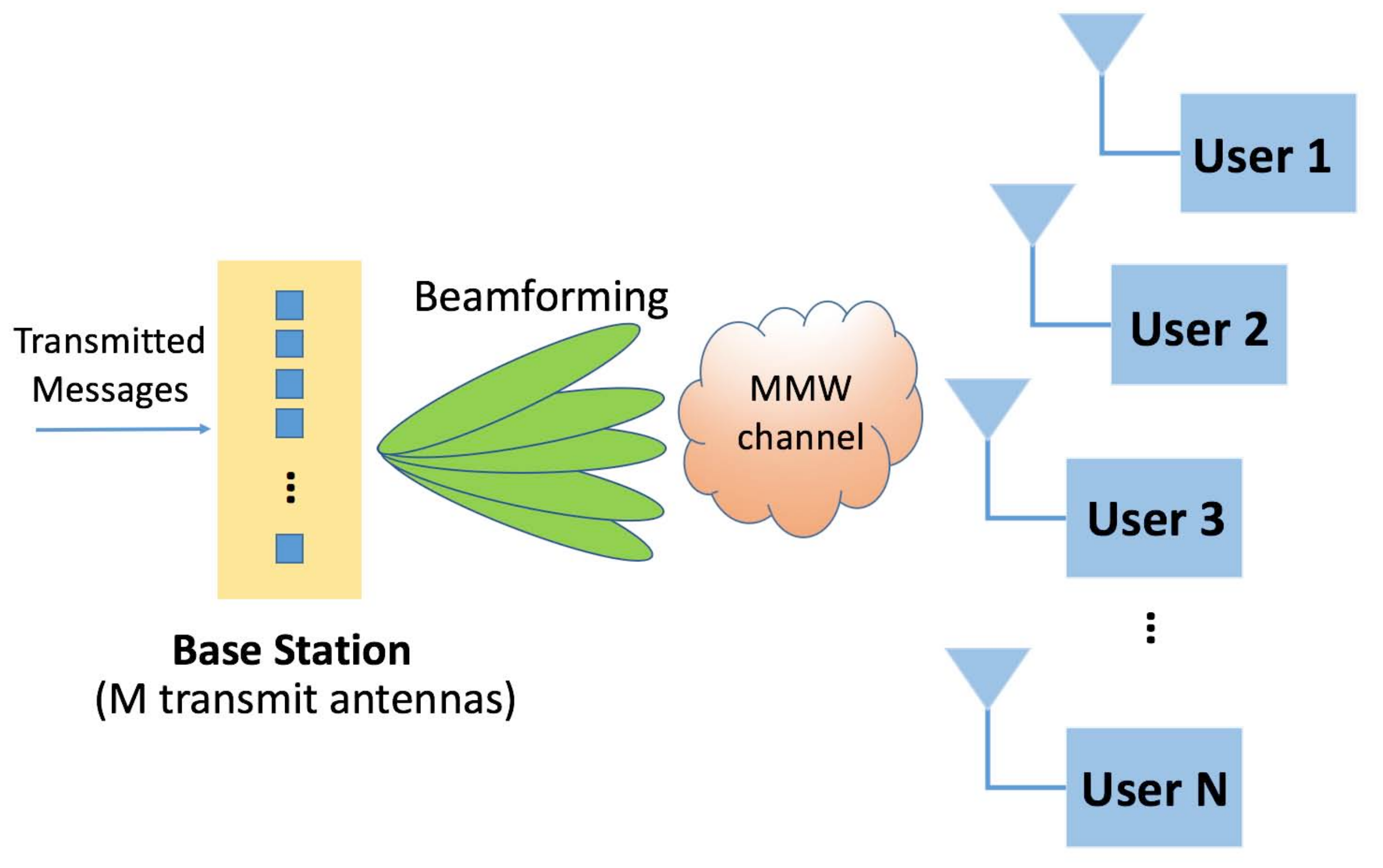}
\caption{MMW multiuser MIMO system model.}
\label{fig_system}
\end{figure}

We use a multiuser MIMO setup with $M$ transmit antennas in a BS and $N$ single-antenna users (see Fig. \ref{fig_system}). At each time slot $t$, the received signal can be described as
    \begin{align}
        \textbf{Y}(t)=\sqrt{\frac{P}{M}}\textbf{H}(t)\textbf{V}(t)\textbf{X}(t)+\textbf{Z}(t),
        \label{equ_Y}
    \end{align}
where $P$ is the total power budget, $\textbf{H}(t)\in \mathcal{C}^{N\times M}$ is the channel matrix, $\textbf{X}(t)\in \mathcal{C}^{M\times 1}$ is the intended message signal, $\textbf{V}(t)\in \mathcal{C}^{M\times M}$ is the precoding matrix, and $\textbf{Z}(t)\in \mathcal{C}^{N\times 1}$ denotes the independent and identically distributed (IID) Gaussian noise matrix. For simplicity, we drop time index $t$ in the following.

The channel $\textbf{H}$ is modeled by
\begin{align}
    \textbf{H} = \sqrt{\frac{k}{k+1}}\textbf{H}_{\text{LOS}}+\sqrt{\frac{1}{k+1}}\textbf{H}_{\text{NLOS}},
    \label{equ_H}
\end{align}
where $\textbf{H}_{\text{LOS}}$ and $\textbf{H}_{\text{NLOS}}$ denote the line-of-sight (LOS) and non-line-of-sight (NLOS) components of the channel. Also, $k$ controls the relative strength of the LOS and NLOS components. It can be seen that $k=0$ represents an NLOS channel, while $k\to\infty$ gives a LOS condition. Also, the NLOS component is assumed to follow complex Gaussian distribution.
\begin{figure}[!t]
\centering
\includegraphics[width=3in]{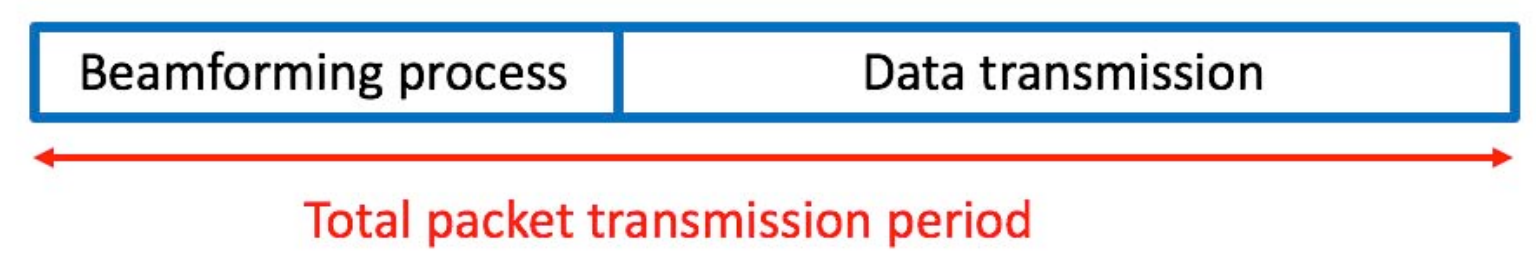}
\caption{Schematic of a packet transmission period.}
\label{fig_timeslot}
\end{figure}

\subsection{Initial Beamforming Procedure}
Conventional beamforming procedure schemes utilize CSI to generate the precoding matrix. However, it is almost impossible to acquire CSI in large scale MMW systems. Instead, we perform codebook-based beamforming, which means selecting a precoding matrix \textbf{V} out of a predefined codebook \textbf{W} at the BS, sending test signal and finally making decisions on transmit beam patterns based on users' feedback about their received metrics. The IA will be finished as soon as a stable control link is established. The time structure for the packet transmission can be seen in Fig. \ref{fig_timeslot}, where part of the packet period is dedicated to design the appropriate beams and the rest is used for data transmission. Thus, we need to find a balance between beamforming time and data transmission period by picking up a suboptimal beamforming approach and using the remaining time for information transmission.
Here, we use a DFT-based codebook [29] defined as 
\begin{align}
   &\textbf{W}  = |w(m,u)| = |e^{-j2\pi(m-1)(u-1)/N_{\text{vec}}}|,\nonumber\\&
        m=1,2,...,M, u=1,2,...,N_{\text{vec}},
\label{equ_V}
\end{align}

where $N_{\text{vec}} \ge M$ is the number of codebook vectors. This codebook can achieve uniform antenna gain in all directions, however, as seen in the following, the proposed algorithm can be implemented for different codebook definitions.

\subsection{Performance Metrics}
As seen in the following, the proposed GA-based algorithm is generic, in the sense that it can be effectively applied for various performance metrics. For the simulations, however, we consider the end-to-end throughput, the service outage-constrained throughput and the service outage probability defined as follows. 

Set $N_{\text{it}}$ to be the maximum possible number of iterations which is decided by designer. Considering the $K$-th iteration round of the algorithm, $K$=1, 2,..., $N_{\text{it}}$, the end-to-end throughput in bit-per-channel-use (bpcu) is defined as
\begin{align}
\label{equ_R}
    & R(K)=(1-\alpha K)\sum_{i=1}^{N}{r_i^K},\nonumber\\&
     r_i^K=\log_2\left(1+\textbf{SINR}_i^K\right).  
\end{align}
Here, $\alpha$ is the relative delay cost for running each iteration of the algorithm which fulfills $\alpha N_{\text{it}}<1$. Also,
\begin{align}
    \textbf{SINR}_i^{K} = \frac{\frac{P}{M}g_{i,i}}{BN_0+\frac{P}{M}\sum_{i\neq j}^{N}g_{i,j}}
\end{align}
is the signal-to-interference-plus-noise ratio (SINR) of user $i$ in  iteration $K$, in which $g_{i,j}$ is the $(i,j)$-th element of the matrix $\textbf{G}^{K}=|\textbf{H}\textbf{V}^{K}|^2$ (and $\textbf{G}$ is referred to as the channel gain throughout the paper), $B$ is the system bandwidth and $N_0$ is the power spectral density of the noise. Thus, $r_i^K$ denotes the achievable rate of user $i$ at the end of the $K$-th iteration of the algorithm. In this way, as opposed to, e.g., [14, Eq. 1] [17, Eq. 43] [19, Eq. 3], [22, Eq. 3], [24, Eq. 5] [26, Eq. 5], we take the algorithm running delay into account. Thus, there is a trade-off between finding the optimal beamforming matrices and reducing the data transmission time slot, and the highest throughput may be achieved by few iterations, i.e., a rough estimation of the optimal beamformer. To simplify the presentations, we set $BN_0=1$. As a result, the power $P$ (in dB, $10\log_{10} P$) denotes the SNR as well.

In different applications, it may be required to serve the users with some minimum required rates, otherwise \emph{service outage} occurs. For this reason, we analyze the end-to-end service outage-constrained throughput defined as

\begin{align}
&\tilde R(K)=(1-\alpha K)\sum_{i=1}^N{r_iU(r_i,\log_2(1+\theta))},\nonumber\\&
U(r_i,\log_2(1+\theta))=\left\{\begin{matrix}
1 & r_i\ge \log_2(1+\theta)\\ 
0 & r_i<\log_2(1+\theta),
\end{matrix}\right.
\label{equ_RN}
\end{align}
where $\log_2(1+\theta)$ is the minimum rate required by the users and $\theta$ represents the minimum required SINR of the users. This is interesting for applications where each user is required to have a minimum rate $\log_2(1+\theta)$. Among our motivations for the service outage-constrained throughput analysis is to highlight the effectiveness of the proposed algorithm in optimizing the non-convex criteria.

Finally, as another performance metric, we study the service outage probability which is defined as
\begin{align}
\phi=\Pr(r_i<\log_2(1+\theta), \forall i).
\label{equ_outage}
\end{align}

\subsection{On the Effect of Power Amplifier}
In multi-antenna systems, when the number of transmit antennas increases, the efficiency of radio-frequency PAs should be taken into account. Here, we consider the state-of-the-art PA efficiency model  [30, Eq. 13], [31, Eq. 3]:
\begin{align}
\label{equ_PA}
    \rho_{\text{cons}} = \frac{ \rho_{\text{max}}^{\mu} } { \epsilon \times \rho_{\text{out}}^{\mu-1} }，
\end{align}
where $\rho_{\text{cons}}$, $\rho_{\text{out}}$, $\rho_{\text{max}}$ refer to the consumed power, output power and maximum output power of the PA, respectively. Also, $ \epsilon \in [0,1]$ is the power efficiency and $\mu \in [0,1]$ is a parameter which depends on the PA classes. Note that setting $\epsilon=1$, $P_{\text{max}}=\infty$ and $\mu=0$ represents the special case with an ideal PA.

\section{Algorithm Description}
We use a GA-based approach for beam selection during IA beamforming and details are explained in Algorithm 1. The algorithm starts by getting $L$ possible beam selection sets randomly and each of them means a certain beam formed by transmit antennas, i.e., a submatrix of the codebook. During each iteration, we determine the best selection result, named as the \textit{Queen}, based on our objective metrics. For instance, we choose the beamforming matrix with the highest end-to-end throughput if (\ref{equ_R}) is considered as the objective function. Next, we keep the Queen and regenerate $S < L$ matrices around the Queen. This can be done by making small changes to the Queen such as changing a number of columns in the Queen matrix. In the simulations, we replace 10\% of the columns of the Queen by other random columns from the codebook. Finally, during each iteration $L-S-1$ beamforming matrices are selected randomly. After  $N_\text{it}$ iterations, considered by the algorithm designer, the Queen is returned as the beam selection rule in the considered time slot.

\begin{algorithm}[ht]
\caption{GA-based Beam Selection Algorithm}
\label{alg:algorithm}
\begin{algorithmic}

\STATE{In each time slot with instantaneous channel realization $\textbf{H}\in \mathcal{C}^{N\times M}$}, do the followings:
\begin{enumerate}[I.]
    \item Consider $L$, e.g., $L = 10$, sets of beamforming matrices $\textbf{V}_l$, $l = 1$,..., $L$,  randomly selected from the pre-defined codebook $\textbf{W}$. 
    \item For each $\textbf{V}_l$, evaluate the instantaneous value of the objective metric $R_{l}$, $l = 1$,..., $L$, for example end-to-end throughput (\ref{equ_R}).
    \item Selection: Find the best beamforming matrix which results in the best value of the considered metric, named as the \textit{Queen}, e.g., $\textbf{V}_q$ satisfies $ R(\textbf{V}_l) \leq R(\textbf{V}_q)$, $\forall$ $l = 1$,..., $L$ if the end-to-end throughput is the objective function.
    \item $\textbf{V}_1$ $\leftarrow$ $\textbf{V}_q$
    \item Create $S$ $\ll$ $L$, e.g., $S = 5$, beamforming matrices $\textbf{V}^{\text{new}}_{s}$, $s = 1$,..., $S$, around the Queen $\textbf{V}_1$. These sets are generated by making small changes in the Queen.
    \item $\textbf{V}_{s+1}$ $\leftarrow$ $\textbf{V}^{\text{new}}_{s}$, $s = 1$,..., $S$.
    \item Go back to Step II and run for $N_{\text{it}}$ iterations, $N_{\text{it}}$ is a fixed number decided by designer.
\end{enumerate}
\STATE{Return the final Queen as the beam selection rule for the current time slot}.

\end{algorithmic}
\end{algorithm}

As demonstrated, the algorithm is generic in the sense that it is independent of the channel model, objective function or precoding matrix, thus this can be used in different scenarios and as a benchmark for comparison of different IA schemes. Also, our proposed algorithm converges to the (sub)optimal value of the considered metrics using few iterations. Therefore, the proposed algorithm can be useful in IA beamforming where the delay is one of the most important factors.

\section{Simulation Results}
%====================================
% fig_ex: figure of algorithm example
%====================================
\begin{figure}[!t]
\centering
\subfloat[System performance with delay ($\alpha=0.001$)]{\includegraphics[width=3in]{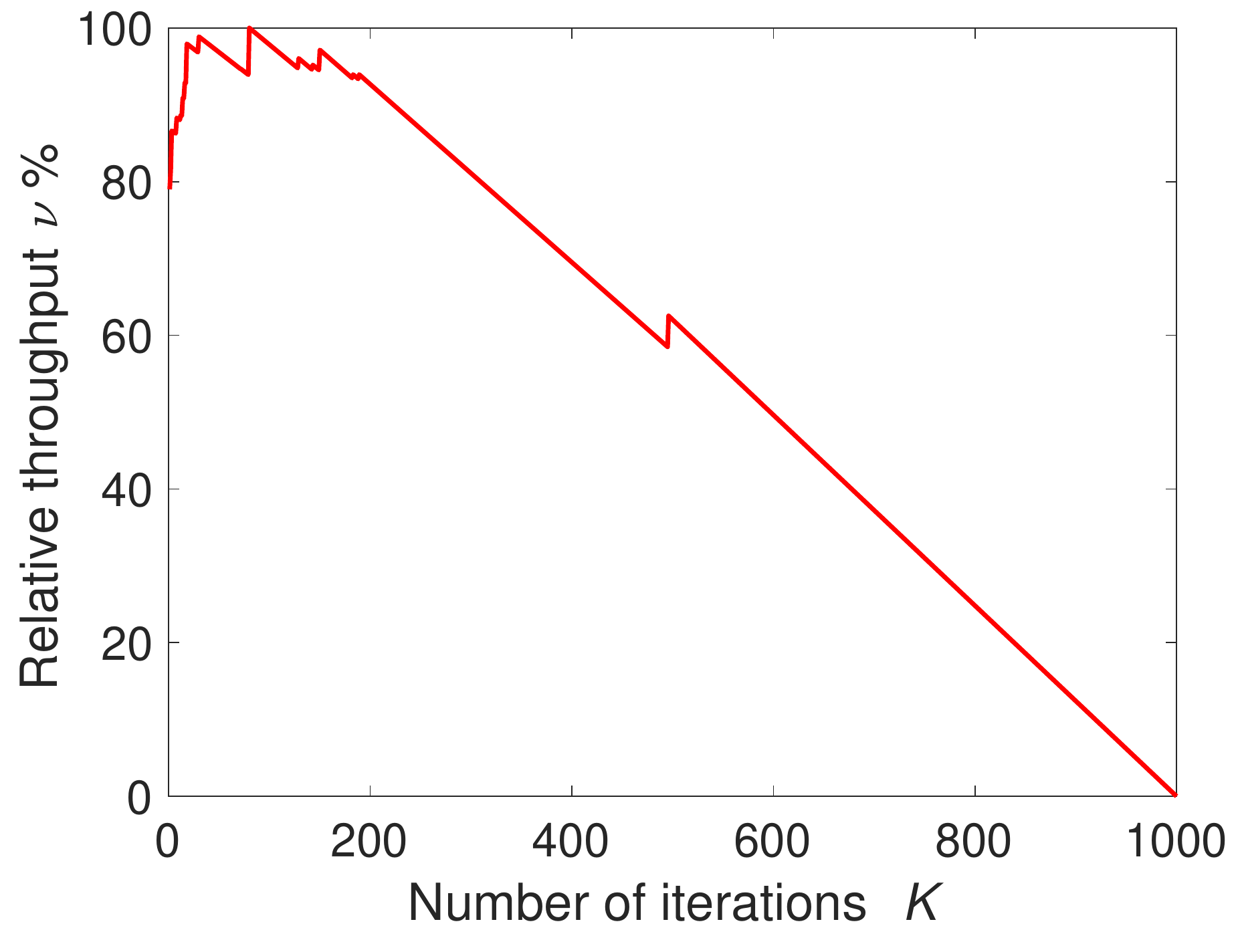}
\label{fig_ex_d}}
\hfil
\subfloat[System performance without delay ($\alpha=0$)]{\includegraphics[width=3in]{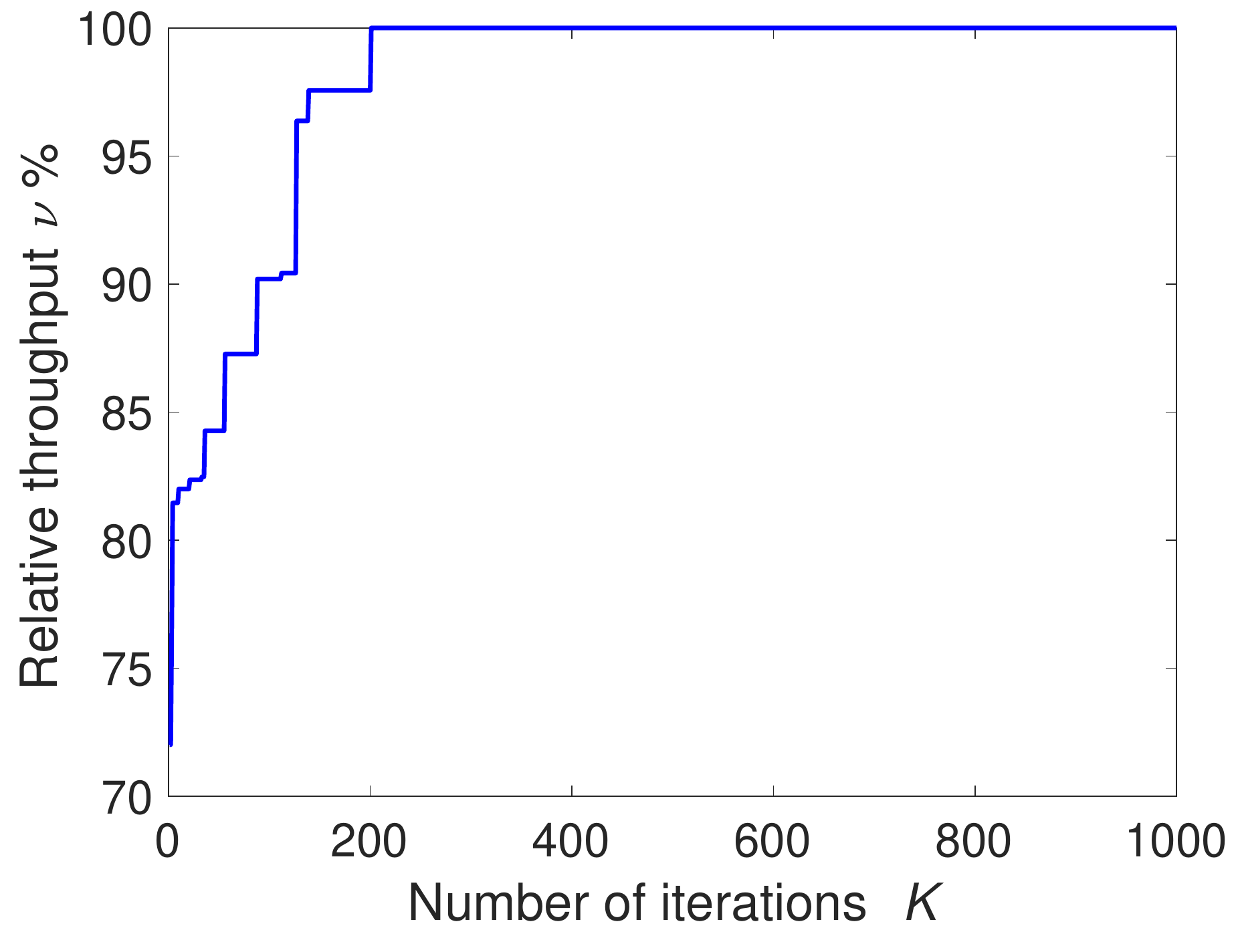}%
\label{fig_ex_nd}}
\caption{Examples of the convergence process of the GA-based beamforming for systems with (subplot a) and without (subplot b) delay cost of the algorithm. $M = 32$, $N = 8$, $N_{\text{vec}}= 128$, $P=10$ dB, $k =0$.}
\label{fig_ex}
\end{figure}

\begin{figure}[!t]
\centering
\includegraphics[width=3in]{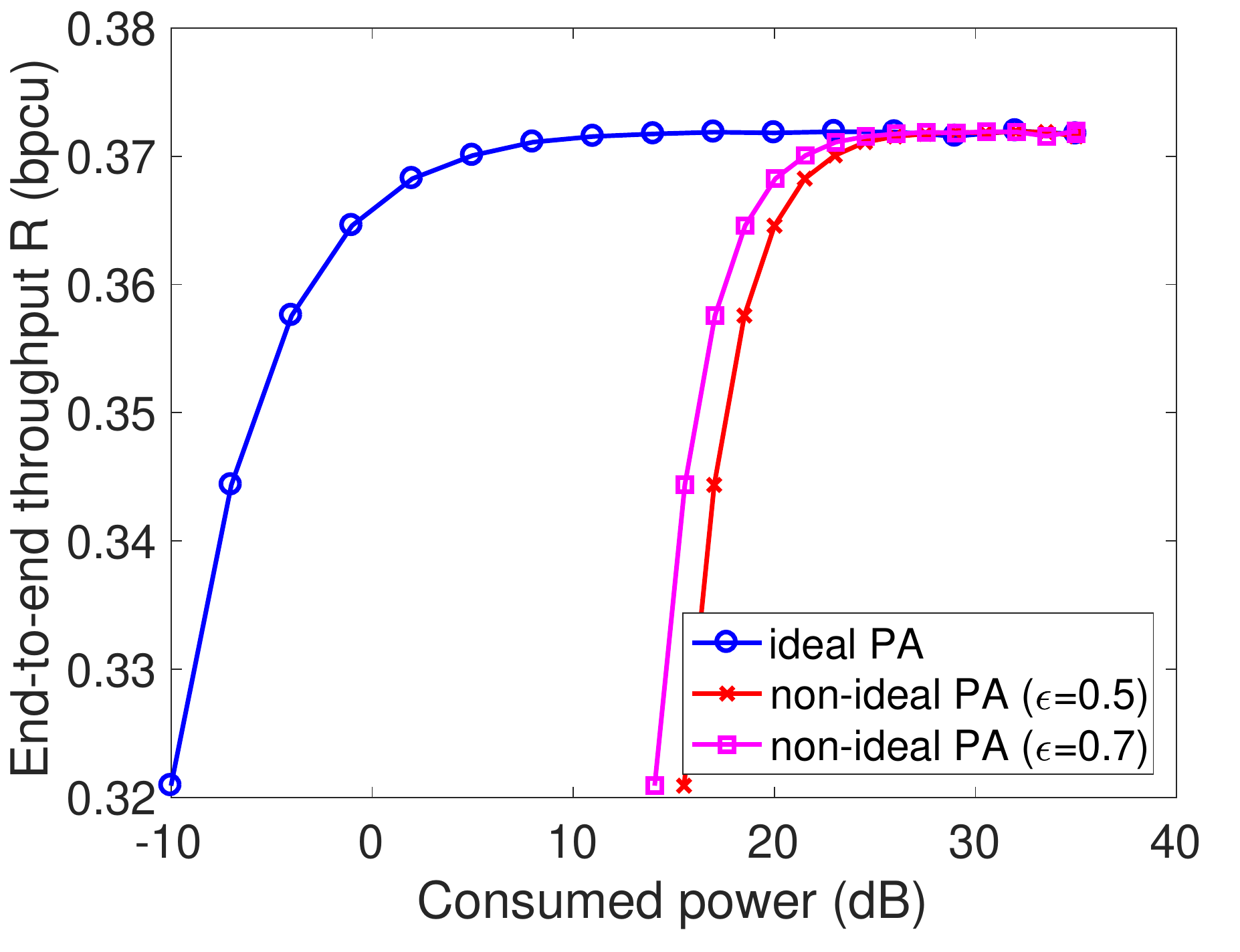}
\caption{The effect of power budget and PAs efficiency on the end-to-end throughput (\ref{equ_R}). $M = 64$, $N = 8$, $k =3$, $N_{\text{vec}}= 128$. $\alpha = 0.001$.}
\label{fig_PA}
\end{figure}

\begin{figure}[!t]
\centering
\includegraphics[width=3in]{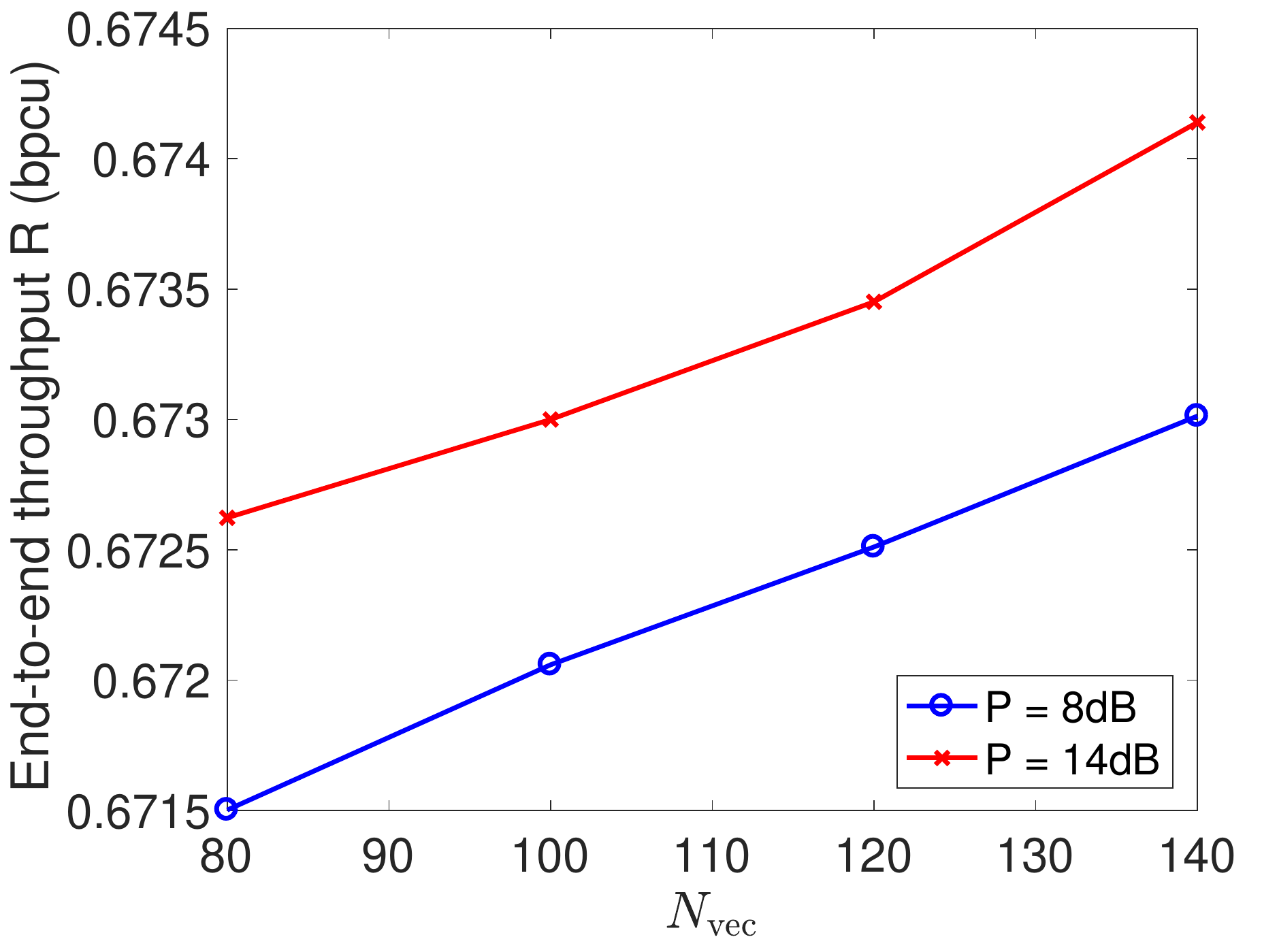}
\caption{The effect of the codebook size on the end-to-end throughput (\ref{equ_R}), $\alpha=0.001$, $M = 32$, $ N = 8$, $k =1$.}
\label{fig_NV}
\end{figure}

\begin{figure}[!t]
\centering
\includegraphics[width=3in]{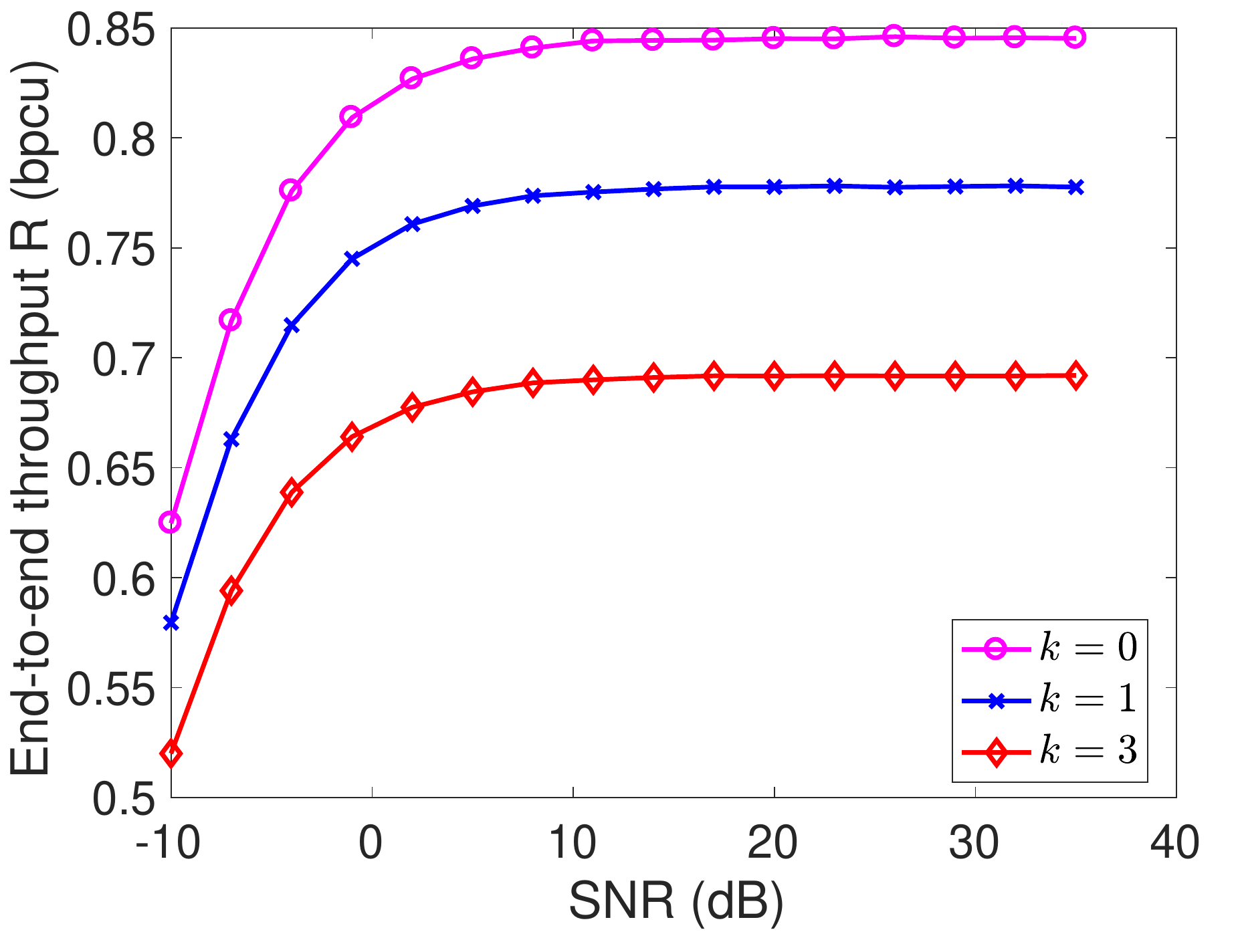}
\caption{The effect of different channel conditions  on the end-to-end throughput (\ref{equ_R}), $M = 32$, $ N = 8$, $\alpha = 0.001$.}
\label{fig_kn}
\end{figure}

\begin{figure}[!t]
\centering
\includegraphics[width=3in]{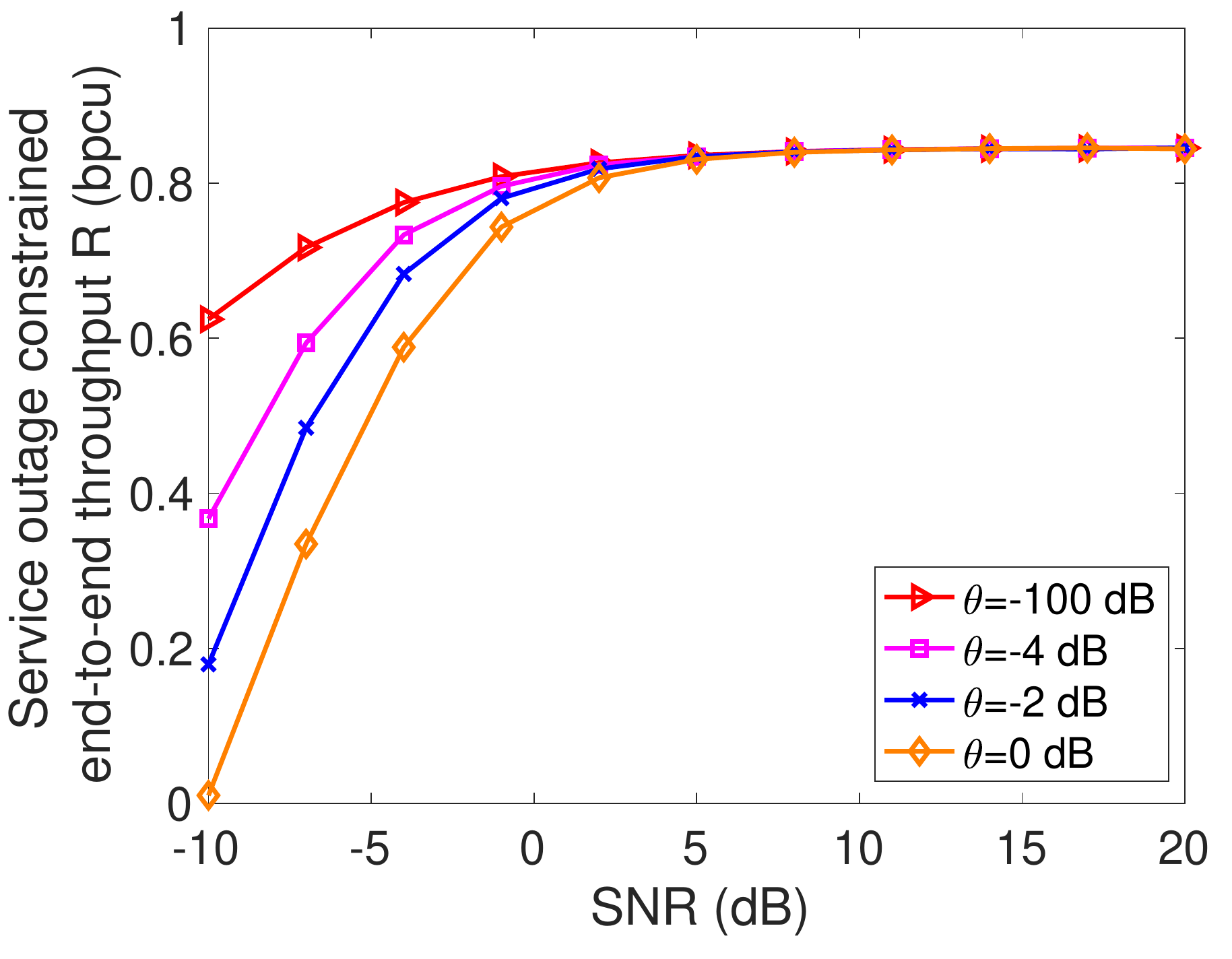}
\caption{Service outage constrained end-to-end throughput with $M = 32$, $ N = 8$,  $k =0$, $N_{\text{vec}}=128$, $\theta = -100, -4,  -2, 0$ dB.}
\label{fig_R_2cases}
\end{figure}

\begin{figure}[!t]
\centering
\includegraphics[width=3in]{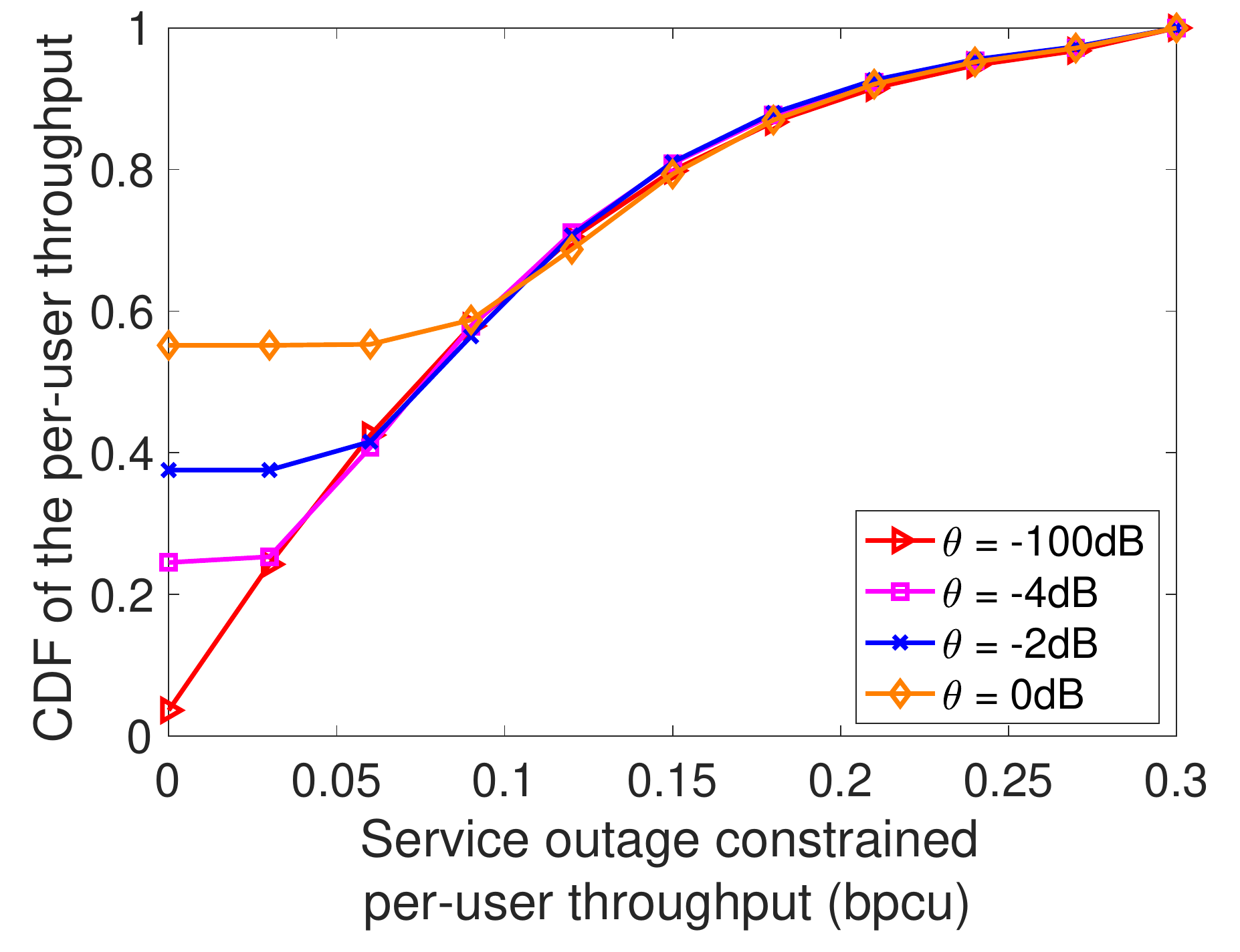}
\caption{CDF of service outage constrained per-user throughput in the cases optimizing (\ref{equ_RN}), $M = 32$, $ N = 8$,  $k =0$, $N_{\text{vec}}=128$, $\theta = -100, -4,  -2, 0$ dB.}
\label{fig_cdf}
\end{figure}

\begin{figure}[!t]
\centering
\includegraphics[width=3in]{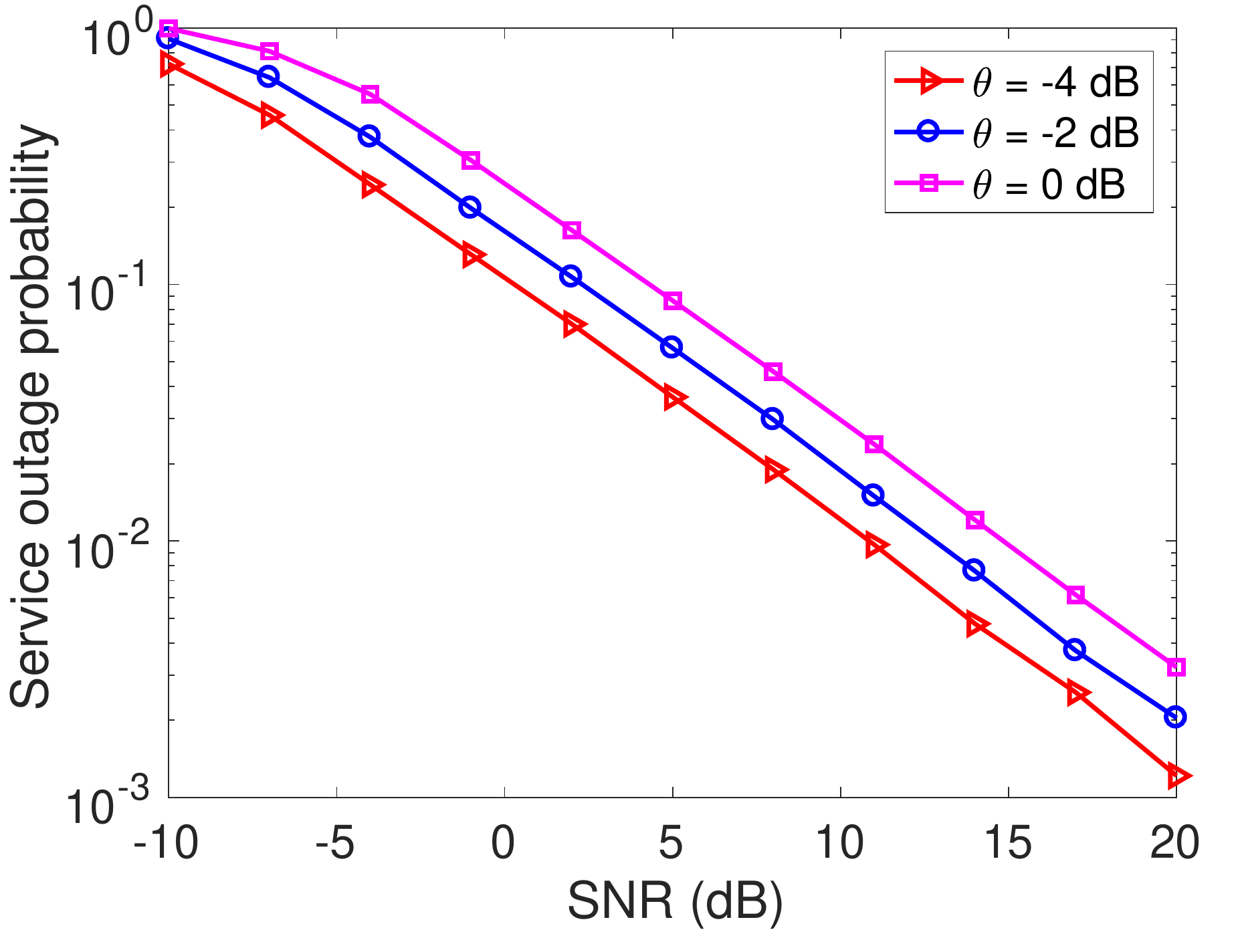}
\caption{Service outage probability in the cases optimizing (\ref{equ_RN}), $M = 32$, $ N = 8$,  $k =0$, $N_{\text{vec}}=128$, $\theta = -4,  -2, 0$ dB.}
\label{fig_outage}
\end{figure}

For the simulation results, we consider the channel model described by (\ref{equ_H}), with $k = 0,1,2,3,10$. We have checked the results for a broad range of parameter settings. However, due to space limits and because they follow the same qualitative behaviors as in Fig. \ref{fig_ex}-\ref{fig_outage}, they are not reported. The initial access beamforming process is performed based on the codebook defined in (\ref{equ_V}). However, one can use the algorithm for different beamforming schemes at the transmitter and receivers. Except for Fig. \ref{fig_ex}  which shows an example of the algorithm procedure by plotting the relative throughput $\nu=\frac{R(K)}{\mathop {\max }\limits_{\forall K}{R(K)}}\%$, for each point in the curves  we run the simulation for $10^4$ different channel realizations of the NLOS component in (\ref{equ_H}). Then, the LOS component of the channel in (\ref{equ_H}) is set to $\textbf{H}_{\text{LOS}}(i,i)=\beta(1+j)$, $\forall i=1$, $\ldots,\min(M,N)$,  $\textbf{H}_{\text{LOS}}(i,k)=\sqrt{\frac{MN-\min(M,N)\beta^2}{\min(M,N)}}(1+j)$, $\forall i\ne k$ or $i=k>\min(M,N)$, $j=\sqrt{-1}$,  which leads to $|\textbf{H}_{\text{LOS}}|^2=1$, $\forall M,N,\beta$. In the simulations we set $\beta=0.2$. In all figures, the GA algorithm is run for sufficiently large number of iterations until no  performance improvement is observed. Then, Tables \ref{table_with}-\ref{table_without} show the average number of required iterations to reach the (sub)optimal solution. Also, we set $L=10, S=5$ in Algorithm 1. Finally, in all figures, except for Fig. \ref{fig_PA}, we consider an ideal PA, i.e., set $P_\text{max}=\infty,\mu=0, \epsilon=1$ in (\ref{equ_PA}). The effect of imperfect PAs is studied in Fig. \ref{fig_PA}. In Figs. \ref{fig_ex}-\ref{fig_kn} and Tables \ref{table_with}-\ref{table_without}, we consider the end-to-end throughput (\ref{equ_R}) as the performance metric. Throughput optimization with a service outage constraint, i.e., (\ref{equ_RN}), is studied in Figs. \ref{fig_R_2cases}-\ref{fig_outage}. Finally, Table \ref{table_N} studies the system performance in the case minimizing the service outage probability (\ref{equ_outage}).

%====================================
% table_NIT: table of needed NIT
%====================================
\newcommand{\tabincell}[2]{\begin{tabular}{@{}#1@{}}#2\end{tabular}}
\begin{table}[!t]
\small
\caption{Average number of required iterations ($\alpha=0.001$)} % title of Table
\vspace{3ex}
\centering % used for centering table
\begin{tabular}{|c|c|c|c|c|c|} % centered columns (2 columns)
\hline %inserts double horizontal lines
$M/N$ & $k=0$ & $k=1$ & $k=2$ & $k=3$ & $k=10$ \\ [0.5ex] % inserts table heading
\hline % inserts single horizontal line
32/8 & 58 & 52 & 47 & 44 & 35\\
\hline
64/8 & 60 & 52 & 46 & 43 & 29\\
\hline %inserts single line
\end{tabular}
\label{table_with} % is used to refer this table in the text
\end{table}

\begin{table}[!t]
\small
\caption{Average number of required iterations ($\alpha=0$)} % title of Table
\vspace{3ex}
\centering % used for centering table
\begin{tabular}{|c|c|c|c|c|c|} % centered columns (2 columns)
\hline %inserts double horizontal lines
 $M/N$ & $k=0$ & $k=1$ & $k=2$ & $k=3$ & $k=10$ \\ [0.5ex] % inserts table heading
\hline % inserts single horizontal line
 32/8 & 495 & 488 & 484 & 483 & 474\\
\hline
 64/8 & 497 & 496 & 491 & 490 & 488\\

\hline %inserts single line
\end{tabular}
\label{table_without} % is used to refer this table in the text
\end{table}
Figures \ref{fig_ex_d} and \ref{fig_ex_nd} show examples of the GA performance in different iterations in the cases with ($\alpha=0.001$) and without costs of running the algorithm ($\alpha=0$), respectively (see (\ref{equ_R})). Here, we set $M = 32$, $N = 8$, $ N_{\text{vec}}= 128$, $P=10$ dB, $k =0$. From Fig. \ref{fig_ex_d} we observe that very few iterations are required to reach the maximum throughput, if the running delay of the algorithm is taken into account. That is, considering the cost of running the algorithm, the maximum throughput is obtained by finding a suboptimal beamforming matrix and leaving the rest of the time slot for data transmission (see Fig. \ref{fig_timeslot}). On the other hand, as the number of iterations increases, the cost of running the algorithm reduces the end-to-end throughput converging to zero at $K=\frac{1}{\alpha}$ (see (\ref{equ_R})). With no cost for running the algorithm, on the other hand, the system performance improves with the number of iterations monotonically (Fig. \ref{fig_ex_nd}). However, the developed algorithm leads to (almost) the same performance as the exhaustive search-based scheme (from Fig. \ref{fig_ex_nd} we can see that $K\to\infty$, enabled by $N_{\text{it}}\to\infty$, represents exhaustive search) with very limited number of iterations (note that with the parameter settings of Fig. \ref{fig_ex}, exhaustive search implies testing in the order of $10^{30}$ possible beamforming matrices). For example, with the parameter settings of Fig. \ref{fig_ex_nd}, our algorithm reaches more than 95\% of the maximum achievable throughput with less than 200 iterations. Furthermore, Tables \ref{table_with} and \ref{table_without} show the average number of iterations that is required in delay-constrained and delay-unconstrained systems to achieve the (sub)optimal system throughput, in the case with $k = 0,1,2,3,10$ and different number of transmit antennas/users. As demonstrated, for both delay-constrained and delay-unconstrained cases the maximum end-to-end throughput is achieved with few iterations of the algorithm. Also, the required number of iterations is almost independent of the channel model and decreases if the cost of running the algorithm is taken into account (Tables \ref{table_with} and \ref{table_without}).

%====================================
% fig_PA: figure of Power amplifier
%====================================

Figure \ref{fig_PA} evaluates the effect of power budget on the end-to-end throughput. Considering $M = 64$, $ N = 8$, $N_{\text{vec}}= 128$, $k =3$, we plot the end-to-end throughput versus the consumed power (see (\ref{equ_PA})). The effect of non-ideal PAs is also considered where we set  $\rho_{\text{max}}=35$ dB, $\mu=0.5$, $\epsilon=0.5, 0.7$. As demonstrated in the figure, the inefficiency of the PA affects the end-to-end throughput remarkably. However, the effect of the PA inefficiency decreases with the SNR. This is intuitively because the effective efficiency of the PAs $\epsilon^\text{effective}=\epsilon(\frac{p_\text{out}}{p_\text{max}})^\mu$ increases with SNR.

%====================================
% fig_NV: figure of codebook vectors
%====================================
In Fig. \ref{fig_NV}, we analyze the effect of the codebook size $N_{\text{vec}}$ on the end-to-end throughput. Here, we set $M = 32$, $ N = 8$,  $k =1$, $P= 8$ dB, $10$ dB. Figure \ref{fig_NV} shows  the end-to-end throughput in a delay-constrained system ($\alpha=0.001$) with different numbers of $N_{\text{vec}}$ in (\ref{equ_V}). As expected, the end-to-end throughput increases (almost) linearly with $N_{\text{vec}}$, because there are more options to select the appropriate beamformer as $N_{\text{vec}}$ increases.

%====================================
% fig_kn: figure of different Kn
%====================================
Figure \ref{fig_kn} shows the effect of the channel condition on the end-to-end throughput. Here, we set $M = 32$, $ N = 8$,  $k =0, 1, 3$, and plot the end-to-end throughput versus the SNR. As seen in the figure, the throughput decreases with $k$, because with the parameter settings of the figure the interference power increases more than the useful signal power as the power of the LOS signal components increase.

%====================================
% fig_R: figure of sum throughput (2 cases)
%====================================

%====================================
% fig_snr: figure of snr threshold (two system)
%====================================

%====================================
% fig_outage: figure of outage probability (two system)
%====================================

Figure \ref{fig_R_2cases} demonstrates the service outage constrained end-to-end throughput (\ref{equ_RN}) for different values of required received SNR thresholds $\theta$ in (\ref{equ_RN}). Also, Fig. \ref{fig_cdf} shows the cumulative distribution function (CDF) of the users achievable rates for different service outage constraints. Here, the results are presented for $N=8, M=32, k =0, N_{\text{vec}}=128$. Finally, Fig. \ref{fig_outage}  studies the service outage probability in the cases optimizing (\ref{equ_RN}). As demonstrated in the Figs. \ref{fig_R_2cases}-\ref{fig_cdf}, the service outage constraint affects the end-to-end and the per-user throughput significantly at low SNRs/severe service outage constraints. For instance, with the parameter settings of Fig. \ref{fig_cdf}, more than 50\% of the users may receive data with rates less than $1$ bpcu, corresponding to $\theta=0$ dB (see Fig. \ref{fig_cdf}, (\ref{equ_RN})). However, the effect of the service outage probability decreases as the SNR increases or $\theta$ decreases (Figs. \ref{fig_R_2cases}-\ref{fig_outage}).

%====================================
% table_N: table of active users
%====================================
\begin{table}[H]
\small
\caption{Average number of served users in Service outage-constrained systems} % title of Table
\vspace{3ex}
\centering % used for centering table
\begin{tabular}{|c|c|c|c|c|c|} % centered columns (2 columns)
\hline %inserts double horizontal lines
$\theta$ (dB) & -100 & -4 & -2 & 0 & 2 \\ [0.5ex] % inserts table heading
\hline % inserts single horizontal line
Case1 & 8.00 & 7.14 & 6.82 & 6.49 & 6.15\\
\hline
Case2 & 8.00 & 7.47 & 7.15 & 6.82 & 6.48\\
\hline %inserts single line
\end{tabular}
\label{table_N} % is used to refer this table in the text
\end{table}

In Table \ref{table_N}, we study the average number of served users in Case 1 and 2 where the service outage constrained throughput (\ref{equ_RN}) and the service outage probability (\ref{equ_outage}) are considered as the optimization metric, respectively. Here, the results are presented for $M = 32$, $ N = 8$,  $k =0$, $N_{\text{vec}}=128$. As expected, compared to Case 1 maximizing (\ref{equ_RN}), the average number of served users increases when the goal is to minimize the service outage probability (Case 2). This is indeed at the cost of some end-to-end throughput loss. Finally, the average number of served users increases as  $\theta$ decreases, i.e., the users minimum required rate decreases.

\section{Conclusion}
We studied the performance of initial access beamforming in delay-constrained networks. Considering delay cost of the beamforming procedure, we evaluated the end-to-end throughput as well as the service outage-constrained system performance under different parameter settings. We developed a GA-based beam selection approach which can reach almost the same throughput as in the exhaustive search-based approach with relatively few iterations. Moreover, the proposed algorithm can be effectively applied for different channel models, performance metrics and beamforming schemes. Therefore, our proposed algorithm is suitable for delay-constrained systems and it can be practically implemented in the future. Then, non-ideal PA affects the system performance remarkably, while its effect decreases at high SNR. Finally, the users' severe service outage constraints affect the end-to-end throughput considerably, while its effect decreases at high SNRs. Comparison between different initial access techniques is an interesting future work on which we are working.


\section*{Acknowledgment}

The research leading to these results received funding from the European Commission H2020 programme under grant agreement $n^{\circ}$671650 (5G PPP mmMAGIC project), and from the Swedish Governmental Agency for Innovation Systems (VINNOVA) within the VINN Excellence Center Chase.

%%%%%%%%%%%%%%%%%%%%%% \IEEEtriggeratref{22}


% \bibliographystyle{IEEEtran}

% \bibliography{ALL}


% \bibliography{IEEEabrv, sample}

%\begin{thebibliography}{1}
%\end{thebibliography}

\end{document}